# Harmonic Representation of Combinations and Partitions

*Michalis Psimopoulos*
*Plasma Group, Physics Department, Imperial College, London SW7 2BZ, UK*
*Email:m.psimopoulos@imperial.ac.uk*

**Abstract.** In the present paper (i) the number of positive integer solutions of the equation $n_1+n_2+n_3+\cdots+n_s=N$ , known also as *Bose states* of a system of particles, is expressed in terms of harmonic functions by the integral

$$\frac{2}{\pi}\int_0^{\pi/2}\left\{\frac{\sin[(N+1)x]}{\sin x}\right\}^s\cos[(s-2)Nx]\,dx=\frac{(N+s-1)!}{N!\,(s-1)!}$$

(ii) the number of positive integer solutions of the equation $n_1+2n_2+3n_3+\cdots+sn_s=s$ representing also the number of energy states of a conservative system of particles as well as the number of *partitions* of the positive integer $s$, is expressed in terms of harmonic functions by the integral

$$p_s=\frac{2}{\pi}\int_0^{\pi/2}\prod_{\kappa=1}^{s}\left\{\frac{\sin[\kappa(s+1)x]}{\sin(\kappa x)}\right\}\cos\left\{\left[\frac{s^2(s+1)}{2}-2s\right]x\right\}\mathrm{d}x$$

## 1. Introduction

The number $B(N,s)$ of integer solutions of the equation

$$n_1+n_2+n_3+\cdots+n_s=N \tag{1}$$

where $n_1\geq 0,\ n_2\geq 0,\ldots,\ n_s\geq 0$, was calculated heuristically by S. Bose who considered the equivalent problem[1] of defining the number of states formed by $N$ particles distributed within $s$ cells. In particular, each state was represented as follows

. . . |. .| . . . |. . | . . .

i.e by $N$ points and by $s$-1 lines separating successive cells. There are $(N+s-1)!$ permutations of the above elements but only permutations where lines are interchanged with points can give a new state. Permutations where only lines or only points are interchanged do not change a state. Therefore

$$B(N,s)=\frac{(N+s-1)!}{N!\,(s-1)!} \tag{2}$$

The same result can be obtained rigorously by using the Darwin-Fowler technique [2]. Consider the generating function

$$F_s(\mathrm{x})=\sum_{n_1=0}^{\infty}\sum_{n_2=0}^{\infty}\ldots\sum_{n_s=0}^{\infty}x^{\mathrm{n_1+n_2+\cdots+n_s}}\qquad;\quad |x|<1 \tag{3}$$

which can be factorized as

$$F_s(x) = \left(\sum_{n_1=0}^{\infty} x^{n_1}\right)\left(\sum_{n_2=0}^{\infty} x^{n_2}\right)\dots\left(\sum_{n_s=0}^{\infty} x^{n_s}\right)$$

$$F_s(x) = (1 + x + x^2 + \cdots)^s = \frac{1}{(1-x)^s} \tag{4}$$

the derivatives of $F_s(x)$ are given by

$$\frac{\partial F_s}{\partial x} = \frac{s}{(1-x)^{s+1}} \quad ; \quad \frac{\partial^2 F_s}{\partial x^2} = \frac{s(s+1)}{(1-x)^{s+2}} \quad ; \dots \quad \frac{\partial^k F_s}{\partial x^k} = \frac{s(s+1)(s+2)\dots(s+k-1)}{(1-x)^{s+k}} \tag{5}$$

and the Taylor expansion of $F_s(x)$ reads

$$F_s(x) = \sum_{k=0}^{\infty} \frac{x^k}{k!}\left(\frac{\partial^k F_s}{\partial x^k}\right)_{x=0} = 1 + sx + \frac{s(s+1)}{2!}x^2 + \cdots + \frac{s(s+1)\dots(s+N-1)}{N!}x^N + \cdots \tag{6}$$

Clearly, the coefficient of $x^N$ of series (6) which in Eq.(3) compiles all $n_1, n_2, \dots, n_s$ that satisfy Eq.(1), can be identified as $B(N,s)$ given by Eq.(2).

On the other hand, in the case of redistribution of $N$ particles between $s$+1 energy levels $0, \varepsilon, 2\varepsilon, 3\varepsilon, \dots, s\varepsilon$ where $N > s$ and $E = s\varepsilon$ is the total energy of the system, conservation of energy implies that the number of particles $n_1, n_2, \dots, n_s$ existing in the levels *1, 2,..,s* respectively and characterizing each state of the system, satisfy the equation

$$n_1 + 2n_2 + 3n_3 + \cdots + sn_s = s \tag{7}$$

Therefore, the number of states of this system can be identified here as the number of integer solutions of the Eq.(7) where $n_1 \geq 0$, $n_2 \geq 0, \dots, n_s \geq 0$. In turn, this number is equal to number of *partitions* $p_s$ of the positive integer $s$ in the sense [3] that $n_1, n_2, \dots, n_s$ represent respectively the number of times that the numbers *1, 2, ...,s* occur in a certain partition of $s$. Note that due to the existence of the zero energy level containing $n_o$ particles where $N - s \leq n_o \leq N - 1$, the sum $n_1 + n_2 + \cdots + n_s = N - n_o$ is not constant in this case so that the number of integer solutions of Eq.(7) depends only on $s$ and not on $N$. If the Darwin-Fowler technique is applied here, the generating function is given by

$$F_s(x) = \sum_{n_1=0}^{\infty} \sum_{n_2=0}^{\infty} \dots \sum_{n_s=0}^{\infty} x^{\mathrm{n_1+2n_2+\dots+sn_s}} \quad ; \quad |x| < 1 \tag{8}$$

which can be factorized as

$$F_s(x) = \left(\sum_{n_1=0}^{\infty} x^{n_1}\right)\left(\sum_{n_2=0}^{\infty} (x^2)^{\mathrm{n_2}}\right) \dots \left(\sum_{n_s=0}^{\infty} (x^s)^{\mathrm{n_s}}\right) \tag{9}$$

and reduces to a *finite Euler* form

$$F_s(x) = \frac{1}{(1-x)(1-x^2)(1-x^3)\dots(1-x^s)} = 1 + x + 2x^2 + 3x^3 + 5x^4 + \dots + p_s x^s + \dots \tag{10}$$

where the coefficient of $x^k$ in the above series represents the partition $p_k$ only if $k \le s$. In particular, the coefficient of $x^s$ of series (10) which in Eq.(8) compiles all $n_1, n_2, \dots, n_s$ that satisfy Eq.(7), can be identified as the number of partitions $p_s$ of the positive integer *s*. However, unlike the first case where $B(N,s)$ was calculated explicitly, no simple formula for $p_s$ has been derived using the above technique.

In the present article $B(N,s)$ and $p_s$ are expressed by a new representation as integrals of harmonic products. The method used is to add characteristic functions of the *Kronecker* type $\delta(m\text{-}n)$ over all integer mesh points of a *s*- dimensional hypercube and then to express $\delta(m\text{-}n)$ in terms of integrals of orthogonal harmonics. Summing up the harmonics, we obtain kernels of the form $\{\sin[N+1)x]/\sin x\}^s$ for the integral representation of $B(N,s)$ and $\prod_{\kappa=1}^{s} \{\sin[\kappa(s+1)x]/\sin(\kappa x)\}$ for the integral representation of $p_s$. Finally, some formula may be repeated if necessary in the text of various sections in order keep these sections autonomous.

## 2. Harmonic representation of combinations

From the geometrical point of view, Eq.(1) represents an hyperplane in a *s*-dimensional cartesian space $[n_1, n_2, \dots, n_s]$ cutting the axis at $n_1 = N;\ n_2 = N; \dots;\ n_s = N$ respectively. Therefore, we can express formally the number $B(N,s)$ of integer solutions of Eq.(1) as

$$B(N,s) = \sum_{n_1=0}^{N} \sum_{n_2=0}^{N} \dots \sum_{n_s=0}^{N} \delta(n_1 + n_2 + \cdots + n_s - N) \tag{11}$$

where

$$\delta(m-n) = \begin{cases} 1 & m = n \\ 0 & m \neq n \end{cases} \tag{12}$$

There are two ways of expressing $\delta(m-n)$ in terms of orthogonal harmonic functions

I. $$\delta(m-n) = \frac{2}{\pi}\int_0^{\pi} \sin(mx)\sin(nx)dx \tag{13}$$

so that by changing variable $x \rightarrow 2x$, Eq.(11) reads

$$B(N,s) = \frac{4}{\pi}\int_0^{\pi/2} \sin(2Nx)\left\{\sum_{n_1=0}^{N} \sum_{n_2=0}^{N} \dots \sum_{n_s=0}^{N} \sin[(n_1+n_2+\cdots+n_s)2x]\right\}dx \tag{14}$$

Sum over $n_1$

$$\sum_{n_1=0}^{N} \sin[(n_1+n_2+\cdots+n_s)2x] = \frac{\sin[(N+1)x]}{\sin x}\sin[(n_2+n_3+\cdots+n_s)2x+Nx] \tag{15}$$

Sum over $n_1, n_2$

$$\sum_{n_1=0}^{N} \sum_{n_2=0}^{N} \sin[(n_1+n_2+\cdots+n_s)2x] = \frac{\sin[(N+1)x]}{\sin x}\sum_{n_2=0}^{N} \sin[(n_2+n_3+\cdots+n_s)2x+Nx]$$

$$= \frac{\sin^2[(N+1)x]}{\sin^2 x}\sin[(n_3+n_4+\cdots+n_s)2x+2Nx] \tag{16}$$

……………………………………………………………………………………………

Sum over $n_1, n_2, \dots, n_s$

$$\sum_{n_1=0}^{N} \sum_{n_2=0}^{N} \dots \sum_{n_s=0}^{N} \sin[(n_1+n_2+\cdots+n_s)2x] = \frac{\sin^s[(N+1)x]}{\sin^s x}\sin(Nsx) \tag{17}$$

Substituting the above result into Eq.(14) we get

$$B(N,s) = \frac{4}{\pi}\int_0^{\pi/2} \left\{\frac{\sin[(N+1)x]}{\sin x}\right\}^s \sin(2Nx)\sin(Nsx)\,dx \tag{18}$$

II. $\delta(m-n)=\frac{2}{\pi}\int_0^{\pi}\cos(mx)\cos(nx)dx$ (19)

so that by changing variable $x\rightarrow 2x$, Eq.(11) reads

$$B(N,s)=\frac{4}{\pi}\int_0^{\pi/2}\cos(2Nx)\left\{\sum_{n_1=0}^{\mathrm{N}}\sum_{n_2=0}^{\mathrm{N}}\dots\sum_{n_s=0}^{\mathrm{N}}\cos[(n_1+n_2+\cdots+n_s)2x]\right\}dx \tag{20}$$

Sum over $n_1$

$$\sum_{n_1=0}^{\mathrm{N}}\cos[(n_1+n_2+\cdots+n_s)2x]=\frac{\sin[(N+1)x]}{\sin x}\cos[(n_2+n_3+\cdots+n_s)2x+Nx] \tag{21}$$

Sum over $n_1$, $n_2$

$$\sum_{n_1=0}^{\mathrm{N}}\sum_{n_2=0}^{\mathrm{N}}\cos[(n_1+n_2+\cdots+n_s)2x]=\frac{\sin[(N+1)x]}{\sin x}\sum_{n_2=0}^{\mathrm{N}}\cos[(n_2+n_3+\cdots+n_s)2x+Nx]$$

$$=\frac{\sin^2[(N+1)x]}{\sin^2 x}\cos[(n_3+n_4+\cdots+n_s)2x+2Nx] \tag{22}$$

…………………………………………………………………………………………

Sum over $n_1, n_2,\dots, n_s$

$$\sum_{n_1=0}^{\mathrm{N}}\sum_{n_2=0}^{\mathrm{N}}\dots\sum_{n_s=0}^{\mathrm{N}}[\cos(n_1+n_2+\cdots+n_s)2x]=\frac{\sin^{s}[(N+1)x]}{\sin^{s}x}\cos(Nsx) \tag{23}$$

Substituting the above result into Eq.(20) we get

$$B(N,s)=\frac{4}{\pi}\int_0^{\pi/2}\left\{\frac{\sin[(N+1)x]}{\sin x}\right\}^s\cos(2Nx)\cos(Nsx)\,dx \tag{24}$$

Adding Eqs.(18, 24) and taking *B*(*N,s*) from Eq.(2) we obtain

$$\frac{2}{\pi}\int_0^{\pi/2}\left\{\frac{\sin[(N+1)x]}{\sin x}\right\}^s\cos[(s-2)Nx]\,dx=\frac{(N+s-1)!}{N!\,(s-1)!} \tag{25}$$

In the special case *N=s* we have

$$\frac{2}{\pi}\int_0^{\pi/2}\left\{\frac{\sin[(s+1)x]}{\sin x}\right\}^s\cos[(s^2-2s)x]\,dx=\frac{(2s-1)!}{s!\,(s-1)!} \tag{26}$$

As we shall see in next section, the harmonic representation of partitions is an extension of formula (26).

## 3. Harmonic representation of partitions

From the geometrical point of view, Eq.(7) represents an hyperplane in a s-dimensional cartesian space $[n_1, n_2, \ldots, n_s]$ cutting each axis at $n_1 = s$ ; $n_2 = s/2$ ; $n_3 = s/3$ ;...; $n_s = 1$ respectively. We can express formally the number $p_s$ of integer solutions of Eq.(7), which is identical to the number of partitions of the integer $s$, as

$$p_s = \sum_{n_1=0}^{s} \sum_{n_2=0}^{s} \ldots \sum_{n_s=0}^{s} \delta(n_1 + 2n_2 + 3n_3 + \cdots + sn_s - s) \tag{27}$$

where

$$\delta(m-n) = \begin{cases} 1 & m = n \\ 0 & m \neq n \end{cases} \tag{28}$$

Since the upper limit of all sums in Eq.(27) is $s$, the characteristic function $\delta$ spans all integer mesh points of the hypercube $0 \leq n_1 \leq s$ ; $0 \leq n_2 \leq s$ ; ...; $0 \leq n_s \leq s$. This is legitimate because the positive part of the hyperplane defined by Eq.(7) is contained within the hypercube and therefore all the points of this hyperplane that are contributing to $p_s$ will be counted by the sums of Eq.(27).

There are two ways of expressing $\delta(m-n)$ in terms of orthogonal harmonic functions

I. $$\delta(m-n) = \frac{2}{\pi}\int_0^{\pi} \sin(mx)\sin(nx)dx \tag{29}$$

so that by changing variable $x \to 2x$, Eq.(27) reads

$$p_s = \frac{4}{\pi}\int_0^{\pi/2} \sin(2sx) \left\{ \sum_{n_1=0}^{s} \sum_{n_2=0}^{s} \ldots \sum_{n_s=0}^{s} \sin[(n_1 + 2n_2 + 3n_3 + \cdots + sn_s)2x] \right\} dx \tag{30}$$

Sum over $n_1$

$$\sum_{n_1=0}^{s} \sin[(n_1 + 2n_2 + 3n_3 + \cdots + sn_s)2x]$$

$$= \frac{\sin[(s+1)x]}{\sin x} \sin[(2n_2 + 3n_3 + \cdots + sn_s)2x + sx] \tag{31}$$

Sum over $n_1, n_2$

$$\sum_{n_1=0}^{s} \sum_{n_2=0}^{s} \sin[(n_1 + 2n_2 + 3n_3 + \cdots + sn_s)2x]$$

$$= \frac{\sin[(s+1)x]}{\sin x} \sum_{n_2=0}^{s} \sin[(2n_2 + 3n_3 + \cdots + sn_s)2x + sx]$$

$$= \frac{\sin[(s+1)x]\sin[2(s+1)x]}{\sin x \sin(2x)} \sin[(3n_3 + \cdots + sn_s)2x + s(1+2)x] \tag{32}$$

Sum over $n_1, n_2, n_3$

$$\sum_{n_1=0}^{s} \sum_{n_2=0}^{s} \sum_{n_3=0}^{s} \sin[(n_1 + 2n_2 + 3n_3 + 4n_4 + \cdots + sn_s)2x]$$

$$= \frac{\sin[(s+1)x]\sin[2(s+1)x]}{\sin x \sin(2x)} \sum_{n_3=0}^{s} \sin[(3n_3 + 4n_4 + \cdots + sn_s)2x + s(1+2)x]$$

$$= \frac{\sin[(s+1)x]\sin[2(s+1)x]\sin[3(s+1)x]}{\sin x \sin(2x) \sin(3x)} \sin[(4n_4 + \cdots + sn_s)2x + s(1+2+3)x] \tag{33}$$

……………………………………………………………………………………………..

Sum over $n_1, n_2, n_3, \ldots, n_s$

$$\sum_{n_1=0}^{s} \sum_{n_2=0}^{s} \ldots \sum_{n_s=0}^{s} \sin[(n_1 + 2n_2 + 3n_3 + \cdots + sn_s)2x]$$

$$= \frac{\sin[(s+1)x]\sin[2(s+1)x] \ldots \sin[s(s+1)x]}{\sin x \sin(2x) \ldots \sin(sx)} \sin\left[\frac{s^2(s+1)x}{2}\right] \tag{34}$$

Substituting the above result into Eq.(30) we get

$$p_s = \frac{4}{\pi} \int_0^{\pi/2} \frac{\sin[(s+1)x]\sin[2(s+1)x] \ldots \sin[s(s+1)x]}{\sin x \sin(2x) \ldots \sin(sx)} \sin(2sx) \sin\left[\frac{s^2(s+1)x}{2}\right] dx \tag{35}$$

II. $\delta(m-n) = \frac{2}{\pi}\int_0^{\pi} \cos(mx)\cos(nx)dx$ (36)

so that by changing variable $x \rightarrow 2x$, Eq.(27) reads

$$p_s = \frac{4}{\pi}\int_0^{\pi/2} \cos(2sx)\left\{\sum_{n_1=0}^{s}\sum_{n_2=0}^{s}\ldots\sum_{n_s=0}^{s}\cos[(n_1+2n_2+3n_3+\cdots+sn_s)2x]\right\}dx \qquad (37)$$

Sum over $n_1$

$$\sum_{n_1=0}^{s}\cos[(n_1+2n_2+3n_3+\cdots+sn_s)2x]$$

$$=\frac{\sin\,[(\mathrm{s}+1)x]}{\sin x}\cos[(2n_2+3n_3+\cdots+sn_s)2x+sx] \qquad (38)$$

Sum over $n_1, n_2$

$$\sum_{n_1=0}^{s}\sum_{n_2=0}^{s}\cos[(n_1+2n_2+3n_3+\cdots+sn_s)2x]$$

$$=\frac{\sin\,[(\mathrm{s}+1)x]}{\sin x}\sum_{n_2=0}^{s}\cos[(2n_2+3n_3+\cdots+sn_s)2x+sx]$$

$$=\frac{\sin[(\mathrm{s}+1)x]\sin[2(\mathrm{s}+1)x]}{\sin x\sin(2x)}\cos[(3n_3+\cdots+sn_s)2x+s(1+2)x] \qquad (39)$$

Sum over $n_1, n_2, n_3$

$$\sum_{n_1=0}^{s}\sum_{n_2=0}^{s}\sum_{n_3=0}^{s}\cos[(n_1+2n_2+3n_3+\cdots+sn_s)2x]$$

$$=\frac{\sin[(\mathrm{s}+1)x]\sin[2(\mathrm{s}+1)x]}{\sin x\sin(2x)}\sum_{n_3=0}^{s}\cos[(3n_3+4n_4+\cdots+sn_s)2x+s(1+2)x]$$

$$=\frac{\sin[(\mathrm{s}+1)x]\sin[2(\mathrm{s}+1)x]\sin[3(\mathrm{s}+1)x]}{\sin x\sin(2x)\sin(3x)}\cos\,[(4n_4+\cdots+sn_s)2x+s(1+2+3)x] \qquad (40)$$

………………………………………………………………………………………..

Sum over $n_1, n_2, n_3, \ldots, n_s$

$$\sum_{n_1=0}^{s}\sum_{n_2=0}^{s}\ldots\sum_{n_s=0}^{s}\cos\,[(n_1+2n_2+3n_3+\cdots+sn_s)2x]$$

$$=\frac{\sin[(\mathrm{s}+1)x]\sin[2(\mathrm{s}+1)x]\ldots\sin[s(\mathrm{s}+1)x]}{\sin x\sin(2x)\ldots\sin(sx)}\cos\left[\frac{\mathrm{s}^2(\mathrm{s}+1)x}{2}\right] \qquad (41)$$

Substituting the above result into Eq.(37) we get

$$p_s = \frac{4}{\pi}\int_0^{\pi/2} \frac{\sin[(s+1)x]\sin[2(s+1)x]\ldots\sin[s(s+1)x]}{\sin x\sin(2x)\ldots\sin(sx)}\cos(2sx)\cos\left[\frac{s^2(s+1)x}{2}\right]dx \qquad (42)$$

Adding Eqs.(35,42) we obtain:

$$p_s = \frac{2}{\pi}\int_0^{\pi/2} \frac{\sin[(s+1)x]\sin[2(s+1)x]\ldots\sin[s(s+1)x]}{\sin x\sin(2x)\ldots\sin(sx)}\cos\left\{\left[\frac{s^2(s+1)}{2}-2s\right]x\right\}dx \qquad (43)$$

This is an exact integral representation of $p_s$ in terms of harmonic functions. The lower values of $\mu = \frac{s^2(s+1)}{2} - 2s$ are given in the following table:

| **s** | 1 | 2 | 3 | 4 | 5 | 6 | 7 | 8 | 9 | 10 |
|---|---|---|---|---|---|---|---|---|---|---|
| **μ** | -1 | 2 | 12 | 32 | 65 | 114 | 182 | 272 | 387 | 530 |

## 4. Conclusion

In the present article a new method of deriving integral representations of combinations and partitions in terms of harmonic products has been established. This method may be relevant to statistical mechanics and to number theory. The basic results obtained are Eq.(25) for the combinations and Eq.(43) for the partitions.